\documentclass{emulateapj}
\usepackage{natbib}

\newcommand\arcsec{\mbox{$^{\prime\prime}$}}

\newcommand\E[1]{\times10^{#1}}

\begin{document}

\author{Neal Dalal\altaffilmark{1}} 
\affil{Institute for Advanced Study, Einstein Drive, Princeton, NJ 08540}

\author{Charles R.\ Keeton\altaffilmark{1}} 
\affil{Astronomy \& Astrophysics Department, University of Chicago,
5640 S.\ Ellis Ave., Chicago, IL 60637}

\altaffiltext{1}{Hubble Fellow}

\title{(Lack of) lensing constraints on cluster dark matter profiles}

\begin{abstract}
Using stellar dynamics and strong gravitational lensing as 
complementary probes, \citet{sand02,sand03} have recently claimed
strong evidence for shallow dark matter density profiles in several
lensing clusters, which may conflict with predictions of the
Cold Dark Matter paradigm.  However, systematic uncertainties in
the analysis weaken the constraints.  By re-analyzing their data,
we argue that the tight constraints claimed by \citeauthor{sand03}\
were driven by prior assumptions.  Relaxing the assumptions, we
find that no strong constraints may be derived on the dark matter
inner profile from the \citeauthor{sand03}\ data; we find satisfactory
fits (with reasonable parameters) for a wide range of inner slopes
$\rho \propto r^{-\beta}$ with $0<\beta<1.4$.  Useful constraints
on the mass distributions of lensing clusters can still be obtained,
but they require moving beyond mere measurements of lensing critical
radii into the realm of detailed lens modeling.
\end{abstract}

\keywords{gravitational lensing --- dark matter --- galaxies: clusters 
--- galaxies: clusters: individual (Abell 383, MS~2137$-$23)}

\maketitle

\section{Cold Dark Matter or Cored Dark Matter?}

The central density profile of dark matter halos has been the subject
of considerable debate in recent years.  On the theoretical side,
numerous groups have claimed that dissipationless N-body simulations
in the Cold Dark Matter (CDM) model produce universal halo profiles.
Despite vigorous debate over the exact shape of the inner profile,
there is general agreement that predicted CDM halos have central
density cusps, $\rho \propto r^{-\beta}$ with $\beta \sim 1$--$1.5$
\citep[e.g.,][]{nfw,moore,navarro03}.  On the observational side,
similar controversy has raged over the question of whether such
cusps are present in real galaxies, and whether their absence would
challenge the CDM paradigm
\citep[e.g.,][and references therein]{dutton03,simon03}.  
The most difficult issue is accounting for the effects of baryonic
matter, which
contributes directly to the gravitational potential and may also
modify the dark matter distribution \citep[e.g.,][]{blumenthal,
loeb03,el-zant03}.

Strongly lensed arcs in galaxy clusters probe the gravitational
potential on scales ($r\sim50$--$100$ kpc) large enough to avoid
significant baryonic contamination, and hence can provide a relatively
clean test of the predictions of dissipationless N-body simulations.
Several groups have constrained the density profiles of individual
clusters by studying their lensing properties 
\citep{clowe02,athreya02,dahle03},
and in some cases have obtained stringent limits from combined
strong- and weak-lensing analyses \citep[e.g.,][]{kneib03,gavazzi03}.  

Another approach is to disentangle the baryonic and dark matter 
contributions to the net potential on small scales, similar to
studies of galaxy rotation curves.  \citet{kelson02} have used the
velocity dispersion profile of the cD galaxy in the cluster
Abell 2199 together with the kinematics of the cluster members to
decompose the stellar and dark matter components.  They found that
a $\beta=1.5$ dark matter cusp is ruled out by the velocity
dispersion data, while a $\beta=1$ cusp is difficult to reconcile
with a reasonable mass-to-light ratio for the cD galaxy.
\citet[hereafter S03]{sand02,sand03} have gone a step further and
combined dynamical and strong lensing analyses for six clusters.
Two of their clusters provide particularly tight and interesting
results: inner slopes of $\beta=0.57\pm0.11$ for MS~2137$-$23 and
$\beta=0.38\pm0.06$ for Abell 383, which strongly conflict with the
steep cusps predicted for CDM.  Such a conflict on cluster scales
may be more troubling than similar conflicts on galaxy scales, as
there are fewer proposed processes that are capable of disrupting
dark matter cusps in clusters. 

\begin{figure}
\plotone{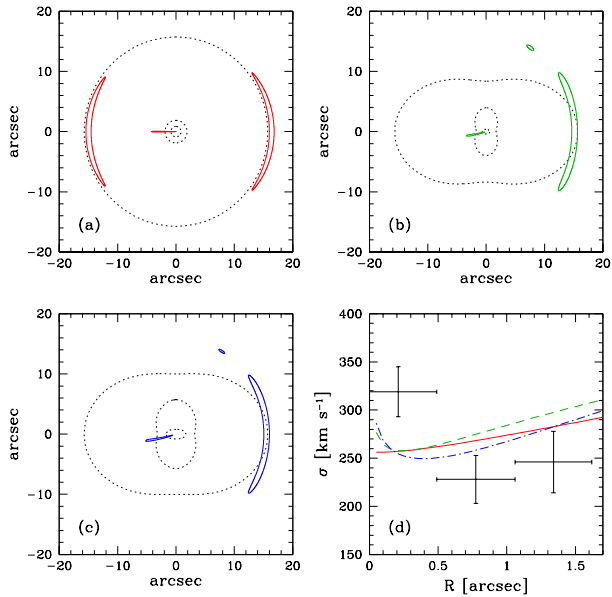}
\caption{(a--c) Representative model arcs for Abell 383.  The model
parameters are listed in Table~\ref{tab1}.  The solid curves show the
arcs, while the dotted curves show the lensing critical curves.
Model a is the best-fit model from S03; the second tangential arc is
an artifact of the assumed spherical symmetry.  In models b and c,
the arclet near the tangential arc is a predicted counter-image of
the radial arc.  (d) Velocity dispersion profiles for models a (solid
line), b (dashed line), and c (dash-dot line), compared with the data
from S03.
\label{counterexample}}
\end{figure}

\begin{deluxetable}{lccc}
\tablecaption{Sample Model Parameters\label{tab1}}
\tablehead{\colhead{Panel} & \colhead{a} & \colhead{b} & \colhead{c}}
\startdata
$M_g$ [$M_\odot$] & $1.8\E{12}$    & $10^{12}$      & $1.2\E{12}$ \\
$r_{\rm hl}$      & $13.75\arcsec$ & $13.75\arcsec$ & $13.75\arcsec$ \\
$e_g$             & 0              & 0.2            & 0.2 \\
$\beta_g$         & 2              & 2.2            & 2.2 \\
$\kappa_g$        & $1.3\E{-2}$    & $3.8\E{-3}$    & $4.2\E{-3}$ \\
$M_h$ [$M_\odot]$ & $4.6\E{15}$    & $1.9\E{15}$    & $4.9\E{14}$ \\
$c_{\rm vir}$     & 9.6            & 4.7            & 24 \\
$r_h$             & $127\arcsec$   & $194\arcsec$   & $24\arcsec$ \\
$e_h$             & 0              & 0.2            & 0.15 \\
$\beta_h$         & 0.4            & 1              & 0 \\
$\kappa_h$        & 0.576          & 0.166          & 1.33
\enddata
\tablecomments{Parameters for the models of Abell 383 shown in
Figure~\ref{counterexample}.  We use $z_l=0.189$, $z_s=1.0$ in a
flat $\Lambda$CDM cosmology with $\Omega_M=0.3$ and $h=0.7$.
Quoted halo masses are virial masses, where the virial overdensity
is approximated using the fitting formula of \citet{bryan98}.
Concentrations are defined by $c_{\rm vir}=r_{\rm vir}/r_h$.
}
\end{deluxetable}

The S03 results are somewhat puzzling, however, as illustrated in
Figure~\ref{counterexample}.  This figure shows velocity dispersion
profiles and representative lensed arcs for three models of Abell 383:
the best-fit model found by S03, a model with a $\beta=1$ dark matter
cusp, and a model with $\beta=0$.  While the fit with a
$\beta=1$ cusp is by no means perfect, it is clearly not ruled out
at the 10-$\sigma$ level (as claimed by S03).  Moreover, it is not
significantly worse than the best-fit model presented by S03.

In this paper we study the origin of this discrepancy.  We argue that
the tight constraints claimed by S03 result from assumptions made in
their modeling.  In particular, they assumed spherical symmetry and
adopted particular values for the scale radius at which the dark
matter density profile transitions from $r^{-\beta}$ to $r^{-3}$.
Relaxing these assumptions, we find that no interesting constraints
on the dark matter profile can be derived from the S03 data alone.
However, more detailed modeling of the lensing properties of these
clusters may yield stronger constraints.

After the completion of this paper, we became aware of independent 
work by \citet{bartelmann03}, who reach similar conclusions.

\section{Analysis}

\subsection{Data and models}

The observables modeled by S03 are the lensing critical radii
(basically the radii of the tangential and radial arcs) and the
velocity dispersion profiles of the brightest cluster galaxies;
the data are given in their Tables 4--5.  For a given mass model,
the critical radii are determined by projecting the density
distribution and locating the singular points of the lens mapping
\citep[see e.g.][]{sef}, while the velocity dispersion profile is
computed by solving the spherical Jeans equation (see appendix).

We use a generalization of the two-component mass model used by S03.
For the stellar component, we use the $\eta$ model \citep{tremaine94},
\begin{equation}
\rho_{\rm gal} = \frac{(3-\beta_g)M_g}{4\pi r_g^3}\left[\left(\frac{r}
{r_g}\right)^{\beta_g}\left(1+\frac{r}{r_g}\right)^{4-\beta_g}\right]^{-1} .
\end{equation}
S03 mainly used a Jaffe model, which is a particular case of the
$\eta$ model with $\beta_g=2$.  Following S03, we describe the dark
matter with a generalized NFW-type profile, 
\begin{equation}
\rho_{\rm DM}=\frac{M_h}{4\pi r_h^3 f(c_{\rm vir})}\left[\left(\frac{r}
{r_h}\right)^{\beta_h}\left(1+\frac{r}{r_h}\right)^{3-\beta_h}\right]^{-1},
\end{equation}
where $f(c)=\int_0^c x^{2-\beta_h}/(1+x)^{3-\beta_h} dx$.

\subsection{A case study: Abell 383}

In the S03 results Abell 383 provided the strongest evidence that the
dark matter slope is shallower than $r^{-1}$, so we use this cluster
to study important systematic uncertainties in the analysis.  If we
center both the stellar and dark matter models on the observed central
galaxy position, orient them along the observed galaxy position
angle, and fix the stellar model's half-light radius to the observed
galaxy effective radius ($13.75\arcsec$), then the remaining model
parameters are: the galaxy mass, inner slope, and ellipticity
$(M_g, \beta_g, e_g)$; and the halo mass, scale radius, inner slope,
and ellipticity $(M_h, r_h, \beta_h, e_h)$.  S03 used a spherical
Jaffe model for the stellar component ($\beta_g=2$ and $e_g=0$),
and a spherical halo ($e_h=0$) with scale radius $r_h=400$ kpc.
With the same assumptions, we reproduce their constraints on the
dark matter inner slope $\beta_h$.

The surface brightness profile of the central galaxy, shown by
\citet{smith01}, is steeper than expected for a Jaffe model and
so we adopt $\beta_g \approx 2.2$, however we note that this steep
slope has been disputed (R.~Ellis 2003, private communication).  
We find that with this stellar
component the models can accommodate a somewhat steeper dark matter
profile: with a $\beta_g=2$ galaxy the best-fit halo has
$\beta_h \approx 0.38$, while with $\beta_g=2.2$ it shifts to
$\beta_h \approx 0.45$.  The shift of $\Delta\beta_h \lesssim 0.1$
is comparable to various systematic effects considered by S03.
The model with a $\beta_g=2.2$ galaxy fits better than the S03
model (with $\Delta\chi^2 = -1.5$), so henceforth we focus on it.

A larger effect is associated with the dark matter scale radius.  The
S03 constraints on $\beta_h$ depend crucially on their assumption of
$r_h=400$ kpc, because there is a degeneracy between $r_h$ and the dark
matter slope $\beta_h$.  For example, assuming $r_h=200$ kpc would
yield $\beta_h \approx 0.18$, while assuming $r_h=800$ kpc would yield
$\beta_h \approx 0.66$ (for spherical models).  The latter model fits
considerably better than the S03 model ($\Delta\chi^2 = -2.6$), even
though S03 claimed that $\beta_h > 0.55$ was excluded at 99\%
confidence.  For spherical models, larger $\beta_h$ do require
large scale radii $r_h$: for example, the best-fitting model with
$\beta_h=0.8$ has $r_h>1$ Mpc!  Such implausibly large scale radii
(or small concentrations) would suggest that steep cusps are still
disfavored, but that is very different from claiming that steep cusps
are inconsistent with the data.  The lesson is that $r_h$ must be
treated as a free parameter, while keeping in mind that its value
should be checked {\em a posteriori\/} for plausibility.

\begin{figure} 
\plotone{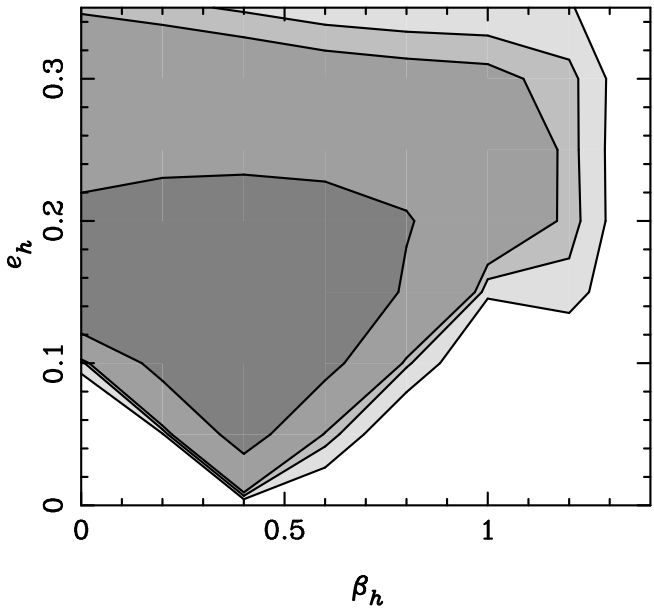} 
\caption{Constraints on dark matter inner slope $\beta_h$ and
ellipticity $e_h$ for Abell 383.  Contours are drawn at the 68, 90,
95 and 99\% confidence levels.
\label{a383e}}
\end{figure}

The most dramatic effects arise upon dropping the assumption of
spherical symmetry.  The ellipticity of the galaxy is not exceedingly
important; we use the observed value $e_g=0.2$ and find that it
does not significantly change the results.  What matters most is
the ellipticity of the halo.  Figure~\ref{a383e} shows likelihood
contours in the plane of $e_h$ and $\beta_h$, optimizing over the
other parameters ($M_g$, $M_h$, and $r_h$).  We find that a broad
range of dark matter slopes are consistent with the data.  In the
limit of spherical symmetry we recover the tight constraints on
$\beta_h$ found by S03, but if we allow even a relatively small
ellipticity we find successful models over the range $0<\beta_h<1.4$.  
Thus, the limits found by S03 appear to be an artifact of their
prior assumptions for the mass model, notably the assumption of
spherical symmetry.  Note that the models we find with steeper inner
slopes ($\beta \approx 1$) have perfectly sensible parameters from a
theoretical standpoint.  For example, model b in Table~\ref{tab1},
with a $\beta_h=1$ cusp, has galaxy mass $M_g=10^{12} M_\odot$, halo
mass $M_h=1.9\E{15} M_\odot$, and a halo scale radius $r_h=610$ kpc,
corresponding to a concentration of $c_{\rm vir}=4.7$.  These halo
parameters are fully consistent with the virial masses and
concentrations expected for hot ($T=7.1$ keV) X-ray luminous
clusters.  (Note that with moderate ellipticities it is possible
to have $\beta_h=1$ without an unphysically large scale radius.)

\subsection{Additional constraints: MS~2137$-$23}

The case of Abell 383 suggests that simply combining the lensing
critical radii with dynamical data cannot strongly constrain the
dark matter slope.  Fortunately, the use of more detailed lensing
data and modeling can provide constraints on the ellipticity that
significantly improve the constraints on the dark matter slope.
To illustrate, we consider MS~2137$-$23.  While S03 used only the
lensing critical radii, \citet{gavazzi03} identified multiple
images of 26 distinct sources in the arcs produced by this cluster,
enabling much more detailed modeling.  Additionally, the X-ray
temperature has been measured to be $T=5.56$ keV \citep{allen01},
which can provide constraints on the halo mass via the $M$--$T$
relation determined by \citet{allen01} for relaxed lensing clusters.
We combine these data with the velocity dispersions from S03, but
we inflate the positional error bars on the multiply imaged knots
reported by \citeauthor{gavazzi03} to $1\arcsec$, and inflate the
error on the X-ray temperature to 1 keV, in order to maximize the
impact of the velocity dispersions on the fit.  (We neglect the
5th image claimed by \citeauthor{gavazzi03}, as its detection
is tentative.)

\begin{figure}
\plotone{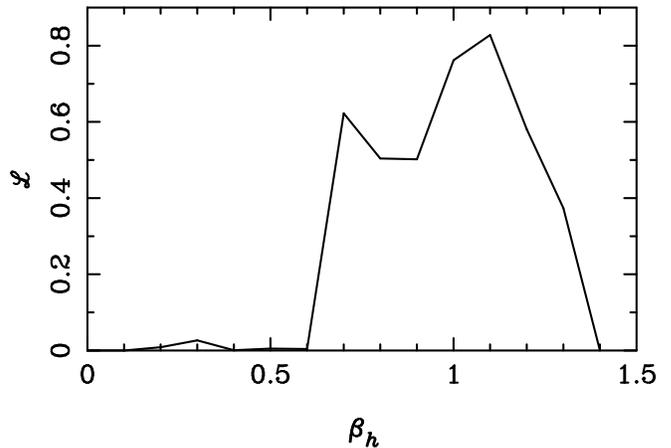}
\caption{Likelihood as a function of the dark matter slope $\beta_h$
in MS~2137$-$23, based on lensing, X-ray temperature, and dynamics.
\label{ms2137}}
\end{figure}

We again use two-component mass models, assuming the galaxy and dark
matter halo to be concentric.  We use the observed values of the
galaxy half-light radius ($5.02\arcsec$), ellipticity ($e_g=0.17$),
and position angle.  If we force the halo to be spherical and to
have scale radius $r_h=400$ kpc we recover the same constraints on
$\beta_h$ as S03.  However, if we allow the dark matter halo
parameters to vary and optimize over them (and also over the galaxy
mass $M_g$), we find the likelihood as a function of the dark matter
slope $\beta_h$ shown in Figure~\ref{ms2137}.  The dark matter
slope is constrained to be $\beta_h = 1\pm0.35$ (95\% confidence).
This is quite consistent with the results found by \citeauthor{gavazzi03},
which perhaps is not too surprising: since we found earlier that
the velocity dispersion data are not terribly restrictive, it is
reasonable that the joint constraints from dynamics and detailed
lens modeling are similar to those obtained from lens modeling alone.
Typical halo virial masses and concentrations obtained were roughly
$M_h \sim 7\E{14} M_\odot$  and $c_{\rm vir}\sim 7$.
Incidentally, we find that mild halo ellipticities ($e_h \sim 0.2$)
and mild misalignment between halo and galaxy
($\Delta\theta \lesssim 10^\circ$) are favored by the fits.

\section{Discussion}

We have argued that the stringent constraints claimed by \citet{sand03}
on the inner slope of the dark matter profiles in clusters ---
$\beta_h=0.38\pm0.06$ for Abell 383, $\beta_h=0.57\pm0.11$ for
MS~2137$-$23 --- are not supported by their data.  We have found
successful models with dark matter slopes well outside these bounds.
The models with steep slopes ($\beta_h \approx 1$) have sensible
parameters: halo virial masses $M_h \sim 10^{15} M_\odot$ and
concentrations $c_{\rm vir} \approx 4$--$5$.

It appears that the tight constraints obtained by S03 on the inner
dark matter profile are artifacts of several simplifying assumptions
in their modeling analysis.  We have shown that the assumption of
spherical symmetry and of a particular value of the halo scale radius
artificially restricts the range of parameters consistent with the
data.  It may seem surprising that small departures from sphericity
(e.g., ellipticities $e \sim 0.1$--$0.2$) generate such striking
differences in the lensing properties of the clusters.  However, as
discussed by \citet{dalal03} and \citet{bartelmann03}, the strong 
lensing cross section for objects with shallow radial profiles 
is highly sensitive to small shear perturbations, arising for
example from ellipticity.  In other words, small changes in
ellipticity elicit large changes in the size of the critical curves
for lenses with shallow profiles.  This extra degree of freedom in
the models thwarts any attempts to derive tight constraints on the
density profiles from the critical radii alone.

The lack of strong constraints does not mean that lensed arcs are
not useful probes of the inner structure of cluster dark matter
halos.  We have also shown that detailed modeling of lensed features
and dynamics does allow interesting constraints to be placed on the
inner potential.  Decomposing the central mass distribution in
MS~2137$-$23 into stellar and dark matter components, we found
constraints on the dark matter inner slope of $\beta = 1\pm 0.35$ at
95\% confidence.

We wouldn't take this measurement too seriously, though.  First, we
have not fully explored the model parameter space --- considering
tidal shear from nearby halos or substructures within the cluster, or
a radially varying ellipticity, could weaken the bounds on $\beta_h$.
Second, it can be argued that decomposing the total mass into stellar
and dark matter components, which is a standard approach, is not ideal.
Because the effects of baryonic mass on the dark matter distribution
are poorly understood, the dark matter profile obtained by subtracting
the baryonic component from the total mass might be very different
from the profile the dark matter would have assumed in the absence
of baryonic interference.  For example, \citet{loeb03} have suggested
that the decomposition procedure could give the appearance of a cored
dark matter distribution, even for proper CDM halos.

Thus, as with galaxy rotation curves, probing the dark matter on
scales dominated by baryons is a perilous exercise.  A much better
test of CDM predictions would be to measure the profile on scales that
are DM dominated.  Fortunately, giant arcs offer such a probe, since
they occur on scales ($\sim 50$--$100$ kpc) beyond the influence of
central galaxies, yet well inside the $\rho \propto r^{-\beta}$
regime expected in clusters.  Simple measurements of lensing critical
radii are not enough, however; detailed lens modeling is required.
Fortunately, the data to support such modeling are available, and
several clusters have already yielded interesting (and in some cases
surprising) measurements of their dark matter distributions
\citep[e.g.,][]{gavazzi03,kneib03}.

\acknowledgments
  We thank many colleagues for their encouragement,
  including Rennan Barkana, Avi Loeb, David Rusin, and Joop Schaye.  
  N.\ D.\ and C.\ R.\ K.\ acknowledge the support of NASA through
  Hubble Fellowship grants HST-HF-01148.01-A and HST-HF-01141.01-A
  awarded by the Space Telescope Science Institute, which is operated
  by the Association of Universities for Research in Astronomy, Inc.,
  for NASA, under contract NAS 5-26555.

\appendix

Here we discuss the calculation of the velocity dispersion and lensing
properties of the mass models defined in \S2.1.  For spherically
symmetric distributions with isotropic velocity dispersion tensors,
the spherical Jeans equations are solved by
\citep{binneytremaine} 
\begin{equation}
\sigma^2(r)=\frac{1}{\rho(r)}\int_r^\infty \rho(r^\prime) 
\frac{G M(r^\prime)}{r^{\prime 2}} dr^\prime,
\end{equation}
where $\rho(r)$ is the stellar density, $G$ is Newton's constant and
$M(r)$ is the total (stellar+DM) mass interior to radius $r$.  This
velocity dispersion is not directly observed; instead the projected,
luminosity-weighted dispersion is measured.  Assuming a constant 
mass-to-light ratio, this becomes
\begin{equation}
\sigma_p^2(R)=\frac{\int_R^\infty\rho(r)\sigma^2(r)\frac{r}{\sqrt{r^2-R^2}}dr}
{\int_R^\infty\rho(r)\frac{r}{\sqrt{r^2-R^2}}dr}=
\frac{2}{\Sigma(R)}\int_R^\infty \rho(r)\frac{G M(r)}{r^2}\sqrt{r^2-R^2}dr,
\end{equation}
where $\Sigma(r)$ is the projected surface density.  Averaging
over a finite radial bin from $R_1$ to $R_2$ gives
\begin{equation}
\sigma_{\rm bin}^2(R_1,R_2) = 
\frac{\int_{R_1}^{R_2} \sigma_p^2(R)\Sigma(R) dR}{\int_{R_1}^{R_2}\Sigma(R)dR}
=\frac{\frac{1}{2}\int_{R_1}^{R_2}\rho(r)GM(r)
\left[F\left(\frac{R_2}{r}\right)-F\left(\frac{R_1}{r}\right)\right]dr}
{\int_{R_1}^{R_2}r\rho(r)
\left[A\left(\frac{R_2}{r}\right)-A\left(\frac{R_1}{r}\right)\right]dr},
\end{equation}
where 
$$
A(x)=\left\{\begin{array}{cl}
\sin^{-1}(x)&x<1\\
\frac{\pi}{2}&x>1
\end{array}\right.\qquad\mbox{and}\qquad F(x)=\left\{
\begin{array}{cl}
x\sqrt{1-x^2}+\sin^{-1}(x)-\frac{\pi}{2}&x<1\\
0&x>1
\end{array}\right.
$$

For the density profiles we have employed, of the form
\begin{equation}
\rho(x=r/r_s) = \frac{\rho_s}{x^\beta(1+x)^{n-\beta}},
\end{equation}
there are no closed-form expressions for the projected surface density
in terms of elementary functions.  The convergence
$\kappa=\Sigma/\Sigma_{\rm crit}$ takes the form
\citep[e.g.,][]{wyithe01}
\begin{equation}
\kappa(u=R/r_s)=2\kappa_s u^{1-\beta}\int_0^{\pi/2}
\frac{(\sin\theta)^{n-2}}{(u+\sin\theta)^{n-\beta}}d\theta,
\end{equation}
where $\kappa_s = \rho_s r_s/\Sigma_{\rm crit}$ and as usual
$\Sigma_{\rm crit}$ is the lensing critical surface density.  For
elliptical surface density profiles with axis ratio $q=1-e$, we use 
$\kappa(\sqrt{x^2+y^2/q^2})$ with
$\kappa_s = \rho_s r_s/(q\Sigma_{\rm crit})$.
\citet{keeton01} discusses how the deflection angle and distortion
tensor may be expressed as one-dimensional integrals over $\kappa$
and its derivatives.  The ellipticities of interest to us are small
enough ($e \sim 0.1$--$0.2$ in the density, even smaller in the
potential) that we assume it is a reasonable approximation to use
the spherical Jeans equations to compute the line-of-sight velocity
dispersion \citep{kochanek94}.


\begin{thebibliography}{27}
\expandafter\ifx\csname natexlab\endcsname\relax\def\natexlab#1{#1}\fi

\bibitem[{{Allen} {et~al.}(2001){Allen}, {Schmidt}, \& {Fabian}}]{allen01}
{Allen}, S.~W., {Schmidt}, R.~W., \& {Fabian}, A.~C. 2001, \mnras, 328, L37

\bibitem[{{Athreya} {et~al.}(2002){Athreya}, {Mellier}, {van Waerbeke},
  {Pell{\' o}}, {Fort}, \& {Dantel-Fort}}]{athreya02}
{Athreya}, R.~M., {Mellier}, Y., {van Waerbeke}, L., {Pell{\' o}}, R., {Fort},
  B., \& {Dantel-Fort}, M. 2002, \aap, 384, 743

\bibitem[{{Bartelmann} \& {Meneghetti}(2003)}]{bartelmann03}
{Bartelmann}, M. \& {Meneghetti}, M. 2003, submitted to \aap, astro-ph/0312011

\bibitem[{{Binney} \& {Tremaine}(1987)}]{binneytremaine}
{Binney}, J. \& {Tremaine}, S. 1987, {Galactic dynamics} (Princeton, NJ,
  Princeton University Press, 1987, 747 p.)

\bibitem[{{Blumenthal} {et~al.}(1986){Blumenthal}, {Faber}, {Flores}, \&
  {Primack}}]{blumenthal}
{Blumenthal}, G.~R., {Faber}, S.~M., {Flores}, R., \& {Primack}, J.~R. 1986,
  \apj, 301, 27

\bibitem[{{Bryan} \& {Norman}(1998)}]{bryan98}
{Bryan}, G.~L. \& {Norman}, M.~L. 1998, \apj, 495, 80

\bibitem[{{Clowe} \& {Schneider}(2002)}]{clowe02}
{Clowe}, D. \& {Schneider}, P. 2002, \aap, 395, 385

\bibitem[{{Dahle}(2003)}]{dahle03}
{Dahle}, H. 2003, ArXiv Astrophysics e-prints, astro-ph/0310549

\bibitem[{{Dalal} {et~al.}(2003){Dalal}, {Holder}, \& {Hennawi}}]{dalal03}
{Dalal}, N., {Holder}, G., \& {Hennawi}, J. 2003, ArXiv Astrophysics e-prints,
  astro-ph/0310306

\bibitem[{{Dutton} {et~al.}(2003){Dutton}, {Courteau}, {Carignan}, \& {de
  Jong}}]{dutton03}
{Dutton}, A.~A., {Courteau}, S., {Carignan}, C., \& {de Jong}, R. 2003, ArXiv
  Astrophysics e-prints, astro-ph/0310001

\bibitem[{{El-Zant} {et~al.}(2003){El-Zant}, {Hoffman}, {Primack}, {Combes}, \&
  {Shlosman}}]{el-zant03}
{El-Zant}, A., {Hoffman}, Y., {Primack}, J., {Combes}, F., \& {Shlosman}, I.
  2003, ArXiv Astrophysics e-prints, astro-ph/0309412

\bibitem[{{Gavazzi} {et~al.}(2003){Gavazzi}, {Fort}, {Mellier}, {Pell{\' o}},
  \& {Dantel-Fort}}]{gavazzi03}
{Gavazzi}, R., {Fort}, B., {Mellier}, Y., {Pell{\' o}}, R., \& {Dantel-Fort},
  M. 2003, \aap, 403, 11

\bibitem[{{Keeton}(2001)}]{keeton01}
{Keeton}, C.~R. 2001, ArXiv Astrophysics e-prints, astro-ph/0102340

\bibitem[{{Kelson} {et~al.}(2002){Kelson}, {Zabludoff}, {Williams}, {Trager},
  {Mulchaey}, \& {Bolte}}]{kelson02}
{Kelson}, D.~D., {Zabludoff}, A.~I., {Williams}, K.~A., {Trager}, S.~C.,
  {Mulchaey}, J.~S., \& {Bolte}, M. 2002, \apj, 576, 720

\bibitem[{{Kneib} {et~al.}(2003){Kneib}, {Hudelot}, {Ellis}, {Treu}, {Smith},
  {Marshall}, {Czoske}, {Smail}, \& {Natarajan}}]{kneib03}
{Kneib}, J., {Hudelot}, P., {Ellis}, R.~S., {Treu}, T., {Smith}, G.~P.,
  {Marshall}, P., {Czoske}, O., {Smail}, I., \& {Natarajan}, P. 2003, \apj, in
  press, astro-ph/0307299

\bibitem[{{Kochanek}(1994)}]{kochanek94}
{Kochanek}, C.~S. 1994, \apj, 436, 56

\bibitem[{{Loeb} \& {Peebles}(2003)}]{loeb03}
{Loeb}, A. \& {Peebles}, P.~J.~E. 2003, \apj, 589, 29

\bibitem[{{Moore} {et~al.}(1998){Moore}, {Governato}, {Quinn}, {Stadel}, \&
  {Lake}}]{moore}
{Moore}, B., {Governato}, F., {Quinn}, T., {Stadel}, J., \& {Lake}, G. 1998,
  \apjl, 499, L5+

\bibitem[{{Navarro} {et~al.}(1997){Navarro}, {Frenk}, \& {White}}]{nfw}
{Navarro}, J.~F., {Frenk}, C.~S., \& {White}, S.~D.~M. 1997, \apj, 490, 493

\bibitem[{{Navarro} {et~al.}(2003){Navarro}, {Hayashi}, {Power}, {Jenkins},
  {Frenk}, {White}, {Springel}, {Stadel}, \& {Quinn}}]{navarro03}
{Navarro}, J.~F., {Hayashi}, E., {Power}, C., {Jenkins}, A., {Frenk}, C.~S.,
  {White}, S.~D.~M., {Springel}, V., {Stadel}, J., \& {Quinn}, T.~R. 2003,
  ArXiv Astrophysics e-prints, astro-ph/0311231

\bibitem[{{Sand} {et~al.}(2002){Sand}, {Treu}, \& {Ellis}}]{sand02}
{Sand}, D.~J., {Treu}, T., \& {Ellis}, R.~S. 2002, \apjl, 574, L129

\bibitem[{{Sand} {et~al.}(2003){Sand}, {Treu}, {Smith}, \& {Ellis}}]{sand03}
{Sand}, D.~J., {Treu}, T., {Smith}, G.~P., \& {Ellis}, R.~S. 2003, ArXiv
  Astrophysics e-prints, astro-ph/0300703 (S03)

\bibitem[{{Schneider} {et~al.}(1992){Schneider}, {Ehlers}, \& {Falco}}]{sef}
{Schneider}, P., {Ehlers}, J., \& {Falco}, E.~E. 1992, {Gravitational Lenses}
  (Gravitational Lenses, XIV, 560 pp.~112 figs..~Springer-Verlag Berlin
  Heidelberg New York.~ Also Astronomy and Astrophysics Library)

\bibitem[{{Simon} {et~al.}(2003){Simon}, {Bolatto}, {Leroy}, \&
  {Blitz}}]{simon03}
{Simon}, J.~D., {Bolatto}, A.~D., {Leroy}, A., \& {Blitz}, L. 2003, \apj, 596,
  957

\bibitem[{{Smith} {et~al.}(2001){Smith}, {Kneib}, {Ebeling}, {Czoske}, \&
  {Smail}}]{smith01}
{Smith}, G.~P., {Kneib}, J., {Ebeling}, H., {Czoske}, O., \& {Smail}, I. 2001,
  \apj, 552, 493

\bibitem[{{Tremaine} {et~al.}(1994){Tremaine}, {Richstone}, {Byun}, {Dressler},
  {Faber}, {Grillmair}, {Kormendy}, \& {Lauer}}]{tremaine94}
{Tremaine}, S., {Richstone}, D.~O., {Byun}, Y., {Dressler}, A., {Faber}, S.~M.,
  {Grillmair}, C., {Kormendy}, J., \& {Lauer}, T.~R. 1994, \aj, 107, 634

\bibitem[{{Wyithe} {et~al.}(2001){Wyithe}, {Turner}, \& {Spergel}}]{wyithe01}
{Wyithe}, J.~S.~B., {Turner}, E.~L., \& {Spergel}, D.~N. 2001, \apj, 555, 504

\end{thebibliography}
\end{document}